\documentclass[12pt]{iopart}

\usepackage{iopams,graphicx}
\begin{document}

\title[Stopping power from SPS to LHC]
{Stopping Power from SPS to LHC energies.}

\def\dmg{$^{1}$}
\def\cunew{$^{2}$}
\def\frank{$^{3}$}

\author{V.Topor Pop\dmg, J.Barrette\dmg, C.Gale\dmg, S.Jeon\dmg and
M.Gyulassy\cunew$^,$\frank}

\address{
\dmg \mbox{McGill University, Montreal, Canada, H3A 2T8}\\
\cunew \mbox{Physics Department Columbia University, New York, New York 10027, USA}\\
\frank \mbox{FIAS, J. W. Goethe Universitat, D-60438,Frankfurt am Main, Germany}\\
}


\begin{abstract}

We investigate the energy dependence of hadron production and 
of stopping power based on HIJING/B\=B v2.0 model calculations.
Pseudorapidity spectra and $p_T$ distributions 
for produced charged particles as well as net baryons (per pair of 
partcipants) and their
rapidity loss are compared to data at RHIC and predictions for
LHC energies are discussed.

\end{abstract}



In previous papers \cite{prc72_top06} we studied 
the possible role of topological baryon junctions \cite{kharzeev_96}
\cite{svance_98}, and the effects of strong color field (SCF) 
in nucleus-nucleus collisions at RHIC energies.
In the framework of HIJING/B\=B v2.0 model,
the new algorithm for junction anti-junction J\=J loops provide a possible 
explanation for baryon/meson anomaly. 
The SCF effects as implemented within our model  
gives a better description of this anomaly. 
At LHC energies, due to higher initial energy density    
(or temperature) we expect an increase of 
the mean value of the string tension ($\kappa$) \cite{top07_proc}.

The day 1 measurements at the LHC will include 
results on multiplicity distributions 
with important consequences for our understanding 
of matter produced in the collisions \cite{nestor_00},\cite{ursw_07}. 
From our model calculations one expects 
dN$^{\rm ch}_{\rm PbPb}$/d$\eta$ $\approx$ 3500 at $\eta =0$
in central (0-5 \%) Pb +Pb collision. 
This correspond to $\approx$ 17.5 produced charged hadrons 
per participant pair. 
These values are higher than those obtained 
by requiring that both limiting fragmentation and the trapezoidal shape
of the pseudo-rapidity distribution persist at the LHC \cite{ursw_07}.
Our model predicts a characteristic violations of the apparently universal
trend, seen up to maximum RHIC energy. In contrast saturation models 
\cite{kharzeev_05} offer a justification 
for the predicted very weak $\sqrt{s_{\rm NN}}$ dependence 
of event multiplicity.


\begin{figure}[h]
\vspace*{-0.2cm}
\begin{center}
\includegraphics*[width=11.8cm,height=5.9cm]{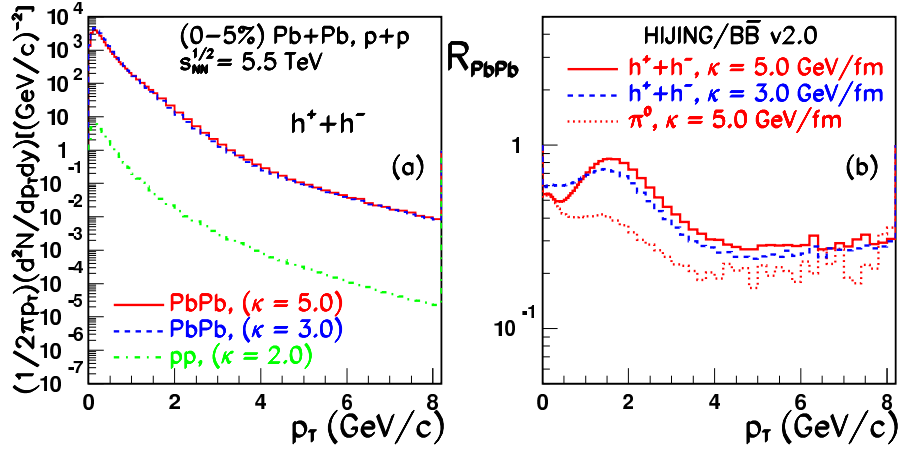}
\end{center}
\vspace*{-0.3cm}
\caption{Left: HIJING/B\=B v2.0 predictions
for $p_T$ spectra at mid-rapidity of total inclusive charged hadrons
for central (0-5\%) Pb+Pb and $p+p$ collisions. Right: 
Predicted nuclear modification 
factors for charged hadrons and for neutral pions.}
\label{chpbpb}
\end{figure}

Figure \ref{chpbpb} presents predictions for $p_T$ spectra at midrapidity
and NMF $R_{\rm PbPb}^{\rm ch}$ of total inclusive charged hadrons
for central (0-5\%) Pb+Pb and $p+p$ collisions at $\sqrt{s_{\rm NN}}$ = 
5.5 TeV. The predicted NMF $R_{\rm PbPb}^{\pi^0}$ of neutral pions 
is also presented.
From our model calculations we conclude that baryon/meson anomaly, will
persist at the LHC with a slight increase for increasing strength
of the chromoelectric field ($\kappa = e_{eff} E$). 
A somewhat higher sensitivity to $\kappa$ is obtained for
NMF of identified particles \cite{top07_proc}. 

\begin{figure}[h]
\vspace*{-0.2cm}
\begin{center}
\includegraphics*[width=11.8cm,height=5.9cm]{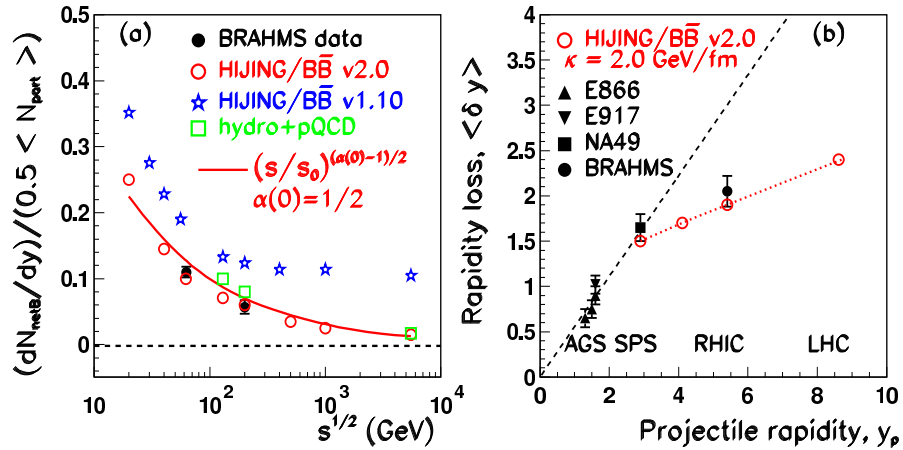}
\end{center}
\vspace*{-0.3cm}
\caption{Left: HIJING/B\=B v2.0 predictions for net-baryon 
(per participant pair) at mid-rapidity as function of $\sqrt{s_{\rm NN}}$. 
Right: Average rapidity loss versus beam rapidity. 
The data and dashed line extrapolation are from ref. \cite{videbaek_06} 
and from BRAHMS \cite{brahms_04}}.
\label{stlhc}
\end{figure}

The net-baryon rapidity distribution measured at RHIC 
is both qualitatively and quantitatively
very different from those at lower energies indicating that a significantly 
different system is formed near mid rapidity \cite{videbaek_06}.
Fig. \ref{stlhc} (left panel) presents the energy dependence  of net-baryon at 
mid-rapidity per participant pair. 
Shown are the results for central (0-5\%) Au+Au collisions, 
which indicate a net decrease with increasing energy.
This picture, corroborated with an increase of the ratio $\bar{p}/p$ to
$\approx$ 1 suggests that the reaction at the LHC is more transparent
in contrast to the situation at lower energy.
For central (0-5\%) Pb+Pb collisions 
at $\sqrt{s_{\rm NN}}$ = 5.5 TeV, our prediction 
for net-baryon per participant pair is  
$\approx$ 0.065 with $N_{\rm part}$ = 398, assuming $\kappa$ = 5 GeV/fm.
Similar values (open squares) are obtained within
pQCD+hydro model \cite{eskola_05}. However, this model 
predicts (Fig. 15 from ref. \cite{eskola_05}) much steeper slopes of 
charged hadron $p_T$ spectra.

In our model the main mechanisms for baryon production are 
quark di-quark ($q-{\rm qq}$) strings fragmentation and
J\=J loops in which baryons are produced approximatively in pairs. 
The energy dependence is $\propto \, (s/s_0)^{-1/4 +\Delta/2}$ 
similar with those predicted in ref. \cite{kharzeev_96} (eq. 11)
with the assumption that J\=J is a dominant mechanisms. 
This dependence is obtained if we choose for the parameters:
$s_0$ = 1 GeV$^2$ the usual parameter  of Regge theory,
$\alpha(0)=1/2$ the reggeon ($M_0^J$) intercept and
$\alpha_{P}(0) = 1 + \Delta$ (where $\Delta \approx 0.01$) for the
pomeron intercept. 
If confirmed, the measurements at LHC energies will
help us to determine better these values.
In contrast, results from HIJING/B\=B v1.10 model 
\cite{svance_98} (star symbol) 
give a slow energy dependence with a higher pomeron intercept 
$\alpha_{P}(0) = 1 + 0.08$ and over-estimate the stopping 
in the entire energy region.

Baryon conservation in the reactions can be used to predict rapidity 
loss and the energy loss per baryon.
The results are illustrated in Fig.~\ref{stlhc} (right panel)
for average rapidity loss 
$<\delta y>$ defined as in ref. \cite{prc72_top06}.
The predicted values for RHIC and LHC energies, 
clearly depart from the linear extrapolation
for constant relative rapidity loss \cite{videbaek_06}, which 
seems to be valid  only at lower energies 
($\sqrt{s_{\rm NN}}\, \leq \,$ 20 GeV).


This work was partly supported by the Natural Sciences and Engineering 
Research Council of Canada and by the U. S. DOE 
under Contract No. DE-AC03-76SF00098 and DE-FG02-93ER-40764.
One of us (MG), gratefully acknowledges partial 
support also from FIAS and GSI, Germany.


\section*{References}

\begin{thebibliography}{20}

\bibitem{prc72_top06} Topor Pop V, Gyulassy M, Barrette J, Gale C,
Wang X N and Xu N 2004 {\it Phys. Rev.} C {\bf 70} 064906;
Topor Pop V, Gyulassy M, Barrette J, Gale C 2006 {\it Phys. Rev.} C {\bf 72}
054901

\bibitem{kharzeev_96} Kharzeev D 1996 {\it Phys. Lett.} B {\bf 378} 238

\bibitem{svance_98} Vance S E, Gyulassy M and Wang X N
1998 {\it Phys. Lett.} B {\bf 443} 45; Vance S E and Gyulassy M 
1999 {\it Phys. Rev. Lett.} {\bf 83} 1735 

\bibitem{top07_proc} Topor Pop V, Barrette J, Gale C, Jeon S 
and Gyulassy M , {\it (these proceedings)}

\bibitem{nestor_00} Armesto N and Pajares C 2000 {\it Int. J. Mod. Phys.} A
{\bf 15} 2019

\bibitem{ursw_07} Wiedemann U A 2007 {\it Preprint} hep-ph/0703146

\bibitem{kharzeev_05} Kharzeev D, Levin E and Nardi M 2005 
{\it Nucl. Phys.} A {\bf 747} 609

\bibitem{videbaek_06} Videbaek F 2006 {\it J. Phys. Conf. Ser.} {\bf 50} 134

\bibitem{eskola_05} Eskola K J, Honkanen H, Niemi H, Ruuskanen P V and
Rasanen S S 2005 {\it Phys. Rev.} C {\bf 72} 044904

\bibitem{brahms_04} BRAHMS Collaboration, Bearden I G {\it et al.} 
2004 {\it Phys. Rev. Lett.} {\bf 93} 102301 



\end{thebibliography}
\end{document}